\documentclass[a4paper,fleqn,usenatbib]{mnras}

\usepackage{newtxtext,newtxmath}

\usepackage[T1]{fontenc}
\usepackage{ae,aecompl}

\usepackage{graphicx}   
\usepackage{amsmath}    
\usepackage{amssymb}    

\title[Short title, max. 45 characters]{Self-trapping as the possible beaming mechanism for FRB}

\author[G. Machabeli et al.]{
G. Machabeli,$^{1}$, A. Rogava$^{1}$\thanks{E-mail:
andria.rogava@iliauni.edu.ge} and B. Tevdorashvili$^{1}$
\\
$^{1}$Centre for Theoretical Astrophysics, ITP, Ilia State
University, Tbilisi 0162, Georgia\\}

\pubyear{2018}

\begin{document}
\label{firstpage}
\pagerange{\pageref{firstpage}--\pageref{lastpage}} \maketitle

\begin{abstract}
Mysterious Fast Radio Bursts (FRB), still eluding a rational
explanation, are astronomical radio flashes with durations of
milliseconds. They are thought to be of an extragalactic origin,
with luminosities orders of magnitude larger than any known short
timescale radio transients. Numerous models have been proposed in
order to explain these powerful and brief outbursts but none of them
is commonly accepted, it is not clear which of these scenarios might
account for real FRB. The crucial question that remains unanswered
is: what makes FRB so exceptionally powerful and so exceptionally
rare?! If the bursts are related with something happening with a
star-scale object and its immediate neighborhood, why all detected
FRB events take place in very distant galaxies and not in our own
galaxy!? In this paper we argue that the non-linear phenomenon -
self-trapping - which may provide efficient but rarely occurring
beaming of radio emission towards an observer, coupled with another,
also rare but powerful phenomenon providing the initial radio
emission, may account for the ultra-rare appearance of FRB.
\end{abstract}


\section{Introduction}

Fast Radio Bursts \citep{lor07,tho13,cha17} (hereafter referred as
FRB) are spatially sporadic and temporarily intermittent radio
emission outbursts of mysterious nature, happening throughout the
universe, with duration of milliseconds. On the basis of a few
credible observational arguments (e.g., observed dispersion measures
greater than the maximum expected from the Galaxy, their spatial
distribution mostly off the Galactic plane) it is strongly
ascertained that FRB are most likely of extragalactic origin. This
circumstance would necessarily imply that the radio luminosities of
related astronomical sources are by several orders of magnitude
larger than any previously detected millisecond-scale radio
transient sources \citep{cor16}.

Originally FRB were detected with large radio telescopes, with
localization accuracy of the order of a few arcminutes. Evidently,
localization efforts have been made and were related with the survey
of simultaneous variability of the immediate neighborhood, adjacent
area galaxies \citep{kea16} or possible presence of peculiar field
stars \citep{loe14}. Until recently these systematic and repeated
efforts failed to pinpoint their location, to lead to the detection
of precise sources of FRB or,at least, their host galaxies with a
satisfactory level of accuracy.

However recently, by means of high-time-resolution radio
interferometric observations, allowing direct imaging  of the bursts
{\it per se}, one of these source, FRB 121102, was localized with a
sub-arcsecond accuracy \citep{cha17}. It appeared to be related to a
persistent and faint radio source with non-thermal continuum
spectrum. It also appears to have a very faint, 25-th magnitude,
optical counterpart. Evidently FRB 121102 remains quite exceptional:
so far it turns out to be the \textit{only} known \textit{repeating
FRB} \citep{spi14,spi16,pet15,sch16}. Even if FRB 121102 is unique
member of the 'family' of RRBs, still the repetitive nature of its
bursts, makes less likely different kinds of 'catastrophic'
scenarios, happening with an astronomical object once in a lifetime.
Another, very important and noteworthy aspect of FRB, is that when
it happens no enhancement of the radiation emission in any other
spectral range has ever been detected.

Evidently, there are quite a number of different models of FRB. For
instance, as early as in 2013 \citep{kas13}, in order to explain
four FRB reported in \citep{tho13}, it was suggested that binary
white dwarf mergers could lead to FRB. A birth of a quark star from
a parent neutron star experiencing a quark nova was also suggested
to be an explanation of FRBs \citep{sha16}. It was also suggested
that FRB might be generated by 'cosmic bomb' - a regular pulsar,
otherwise unnoticeable at a cosmological distance, producing a FRB
when its magnetosphere is suddenly ``combed” by a nearby, strong
plasma stream toward the anti-stream direction \citep{zha17}. It was
also argued that a black hole absorbing a neutron star companion on
the the battery phase of the binary, when the black hole interacts
with the neutron star magnetic field could become a source of at
least a subclass of FRB \citep{min15}. However, this mechanism is
expected to produce electromagnetic radiation mainly in the
high-energy (X-rays and/or gamma-rays) range, while FRB are observed
only in the radio range. In another interesting model \citep{fal14}
FRBs are surmised to represent final signals of a supermassive
rotating neutron star: it is supposed that initially they are above
the critical mass for non-rotating models, supported by their rapid
rotation. But magnetic braking constantly reduces their spins, and
at some moment of time these neutron stars start suddenly collapsing
to a black hole, producing a FRB. A somewhat similar model was
suggested in \citep{ful15}: neutron star collapsing as a result of
'sedimentation' of dark matter (dark matter particles sinking to the
center of a neutron star and becoming the same temperature as the
star) within its core. Eventually, black hole is created at the
center of the neutron star, with the collapse leading to the
powerful radio outburst.

Obviously special attention is focused on the repeating FRB 121102
source. Its persistent radio counterpart is believed to have number
density of particles of the order of $N\sim 10^{52}$, energy about
$E_{N} \sim10^{ 48} $erg, and its length-scale of the order of $R
\sim10^{17} $cm. The FRB source is argued to be a nebula heated and
expanded by an intermittent outflow from a peculiar magnetar  a
neutron star powered not by its rotational energy but by its
magnetic energy \citep{bel17}. The peculiarity of the object is
related with its very young age; it is supposed to liberate its
energy  frequently, in giant magnetic flares driven by accelerated
ambipolar diffusion in the neutron star core. The flares would
eventually feed the nebula and produce bright millisecond bursts. In
\citep{viy17} yet another model for repeating FRBs was proposed,
implying the existence of a variable and relativistic
electron-positron beam, being boosted by an impulsive MHD mechanism,
interacting with a plasma cloud at the center of a dwarf galaxy.
According to this model, the interaction leads to the development of
plasma turbulence and creates areas of high electrostatic field -
cavitons - in the cloud. It is argued that as a result short-lived,
bright coherent radiation bursts, FRB, are generated.

Summarizing, we can cite a very recent review paper by J. I. Katz,
where he says 'More than a decade after their discovery,
astronomical Fast Radio Bursts remain enigmatic. They are known to
occur at 'cosmological' distances, implying large energy and
radiated power, extraordinarily high brightness and coherent
emission. Yet their source objects, the means by which energy is
released and their radiation processes remain unknown'.
\citep{kat18}.

We believe that before trying to involve exotic phenomena for the
explanation of FRB it is reasonable to try to explain FRB on the
basis of traditional, well-established physical phenomenon. In this
letter, we argue that a well-known nonlinear optical phenomenon -
self trapping \citep{chi64,chi65} - could serve as an alternative
FRB model. The advantage of the proposed model is in its
self-sustained and autonomous nature: it doesn't require additional
sources of energy and it naturally provides sufficiently high
emission in a narrow spectral range without leading to a
simultaneous radiation outburst in any other spectral ranges. It is
also worthwhile to note that outbursts of similar nature are
observed for certain objects in our galaxy, for instance, gamma-ray
bursts for the Crab Nebula \citep{buh14}. Recently for the
explanation of these powerful bursts the elements of nonlinear
optics \citep{mac15} have been used.

Before coming directly to the core of the problem and the contents
of our model let us note that nonlinear optics is based on the
fundamental principle of self-focusing of a powerful electromagnetic
wave passing through a nonlinear medium. The self-focusing effect is
related with the dependence of the medium dielectric permittivity on
the wave intensity. A good example of a nonlinear medium is a
liquid/plasma/gas which under the influence of a powerful
electromagnetic wave develops coherent orientation of its molecules
along the field. It leads, in its turn, to the anisotropy of the
medium, increase of the electric field and the growth of the
refraction index. In these circumstances the medium behaves as a
focusing lens for incoming electromagnetic waves with transverse
intensity gradient. The details of the mechanism are considered in
the next section of the paper, while discussion of the model and its
implications in the context of FRB phenomenon are given in the final
section.

\section{Main Consideration}

Let us consider a cloud of relativistic electron-proton plasma which
sustains powerful electrostatic Langmuir waves (plasma oscillations)
of electrons relative to heavy ions. Their wavelength can not be
less than Debye radius, which for a relativistic plasma is:
\begin{equation}
R_{D} = c {{\gamma_{p^{3/2}}}/{\omega_{p}}} ,
\end{equation}
where $\omega_{p} = (4 \pi e^{2} n_{p}/m)^{1/2}$ is Langmuir
(plasma) frequency of nonrelativistic plasma, while $\gamma_{p}$ and
$n_{p}$ are Lorentz factor and number density of particles,
respectively, and $c$ is the speed of light.

The spectrum of Langmuir waves  in relativistic plasma has the form
\begin{equation}
\omega = \sqrt{{\omega_{p}^{2}}/{\gamma_{p}^{3}} + 3 {\bf k}^{2}
c^{2}}
\end{equation}
where ${\bf k}$ is the wavenumber vector of the electrostatic wave.
The factor $3$ is related with the spatial isotropy of the medium.
For a magnetized plasma the isotropy is violated and instead of the
factor $3$ in (2) we have $1$.

For Langmuir waves the Debye radius $R_{D}$ is the charge separation
length-scale and is defined by the distance at which electron
density fluctuation can be shifted on the plasma oscillation period
time-scale due to the thermal motion of electrons. It leads to the
polarization of the medium in the Debye volume caused by the
grouping of charged particles. If one knows the value of the Debye
radius and the average distance between the particles ${\langle d
\rangle} = n_{p}^{1/3}$, then it is possible to estimate the number
of dipoles ($N_{d} = (R_{D}/{\langle d \rangle})^{3}$) in the Debye
volume. We assume that the plasma cloud contains a large number of
Debye volumes but only in one of them the radiation is directed
along the line of sight.Further, we suppose that the current of
charged particles is continuous. Relativistic particles rapidly
leave the volume but they are substituted by other, identical
particles.Therefore the onset of the system with Langmuir waves can
be considered quasi stationary. It allows us to use the dipole
approximation.

Let us now suppose that the Debye volume is filled with spatially
aligned dipoles, oriented towards the line of sight. Obviously, it
is an idealization because, in reality, in the Debye volume, only a
part of the dipoles could be aligned with the line of sight.
Furthermore, let us also assume that through the Debye volume an
electromagnetic wave of radio frequency is passing. Let us assume
that its energy is less than the energy of the electrostatic waves,
but nevertheless it is powerful enough to cause the shifting of the
multitude of dipoles.

The polarization vector has the following form:
\begin{equation}
{\bf P} = e N_{D} {\bf r}
\end{equation}
Even when the electric field ${\bf E}(t)$ of the incident wave is
small it still manages to shift charged particles by a small
displacement value ${\bf r}(t)$. The shifting, in its turn, causes
the appearance of the restoring force ${\bf f}(t) = - \eta {\bf
r}(t)$, where $\eta$ ia an analogue of the spring constant in
Hooke's law. However, when the field ${\bf E}(t)$ is not too small
the displacement ${\bf r}(t)$ can be more considerable and the
expression for the restoring force will contains also a second,
nonlinear term:
\begin{equation}
{\bf f}(t) = - \eta {\bf r}(t) - q {\bf r}^{3}(t)
\end{equation}
where $q$ is a constant coefficient. Its value does not have a
decisive role in the framework of the present consideration.

The value of the electron displacement ${\bf r}(t)$ can be
determined from the equation of motion \citep{mac15}. In a
relativistic case it has the following form:
\begin{equation}
m \gamma^{3} {{d^{2} {\bf r}}\over{dt^{2}}} = - m \gamma \Gamma {{d
{\bf r}}\over{dt}} - \eta {\bf r} - q {\bf r}^{3} + e {\bf E}
\end{equation}
where $\gamma$ is corresponding Lorentz factor and where the first
term on the right hand side of the equation describes dissipation,
with $\Gamma$ being the damping rate.

Taking into account the definition (3) of the polarization vector
{\bf P} from (4) we derive:
\begin{equation}
 {{d^{2} {\bf P}}\over{dt^{2}}} + \Gamma {{d {\bf P}}\over{dt}} + \left({{\omega_0}\over{\gamma}} \right)^{2} {\bf P}
 + Q {\bf P}^{3} = {\left({{e^{2} N_D}\over{m \gamma^{3}}} \right)} {\bf E}
\end{equation}
where $Q \equiv q/m e^{2} N_{p}^{3} \gamma^{3}$ and $\omega_0$, in
this case, is the Langmuir oscillation frequency $\omega_{0} =
\omega_{p}$.

We have noted the incident wave's electric field $\bf E$ is supposed
to be large, but still much less than the intensity of the internal
field within the cloud. In this case the nonlinear term in (6) can
be considered to be small and we can solve the equation by means of
the method of successive approximations. In particular, supposing
$\bf P \equiv \bf P_{L} + \bf P_{NL}$, with $\bf P_{L} \gg \bf
P_{NL}$, and neglecting the nonlinear term we obtain:
\begin{equation}
 {{d^{2} {\bf P_{L}}}\over{dt^{2}}} + \Gamma {{d {\bf P_{L}}}\over{dt}} + \left({{\omega_{0}}\over{\gamma}} \right)^{2} {\bf P_{L}}
= {\left({{e^{2} N_D}\over{m \gamma^{3}}} \right)} {\bf E}
\end{equation}

If we further write down the electric field of the wave as
\begin{equation}
{\bf E}(t) = {\bf A} \cos (\omega t)
\end{equation}
then the solution of (7) is found to be:
\begin{equation}
{\bf P_{L}}(t) = \left(\frac{e^{2} N_{D}}{m \gamma^{3}
\sqrt{(\omega^{2} - \omega_{0}^{2})^{2} + 4 \Gamma^{2}
\omega^{2}}}\right) {\bf A} \cos (\omega t + \Phi)
\end{equation}
where $\tan(\Phi) = \Gamma \omega/(\omega^{2} - \omega_{0}^{2})$.

Note that the Langmuir frequency $\omega \gg \omega_{0}$ is the
frequency of radio emission. Therefore, the range of frequencies we
consider is far from the resonance: $| \omega_{0}^{2} - \omega^{2} |
\gg 4 \Gamma^{2}$ and the impact of the dissipation term can be
neglected. The vector of polarization $\bf P $ is related to the
electric field through the polarization of the medium $\mu$ in the
following way: $\bf P = \mu \bf E$. This expression can be written
as:
\begin{equation}
{\bf P}(t) = \mu(\omega) {\bf E}(t)
\end{equation}

After this linear solution is found the equation in the nonlinear
approximation has the following form:
\begin{equation}
  \frac{\partial^{2} {\bf P}_{NL}}{\partial t^{2}} + \omega_{0}^{2} {\bf P}_{NL}
= - \left( \frac{q \mu^{3}(\omega)}{m \gamma^{3} e^{2} N_{D}^{2}}
\right) {\bf E}^3(t)
\end{equation}

Let us rewrite ${\bf E}^{3}(t)$ using trigonometric identity
$cos^{3}(\omega t) = (1/4) [ 3 cos (\omega t) + cos (3 \omega t)]$.
Then on the right hand side of (11) we have two terms, describing
the contribution of the first and the third harmonics. Accordingly
one can write:
\begin{equation}
{\bf P}(t) = \mu(\omega, A) {\bf E}(t)
\end{equation}
with $\mu(\omega, A)$ defined by: $ \mu(\omega, A) =
\mu(\omega){\left[1+ {{3 q \mu^{2}(\omega) A^{2}}\over{4m n^{2}
e^{2} \omega^{2}}} \right]}. $ while $\mu(\omega)$ is determined
from the solution of (7):
\begin{equation}
\mu(\omega) = {{e^{2} N}\over{m \gamma^{3} \omega^{2}}}
\end{equation}
Note that in the serial expansion of $\mu(\omega, A)$ only first
non-vanishing terms are maintained.

The dielectric permittivity of the medium is described by the tensor
$\varepsilon_{ij}(\omega, A)$. The connection between
$\varepsilon_{ij}(\omega, \b E)$ and $\mu_{ij}(\omega, \b E)$
tensors is given by:
\begin{equation}
\varepsilon_{ij}(\omega, \b E) = \delta_{ij} + 4 \pi
\mu_{ij}(\omega, A)
\end{equation}

The induction vector $\bf D = \bf E + 4 \pi \bf P$, where $D_{i} =
\varepsilon_{ij}(\omega, E) E_{j}$. Subsequently, taking into
account(12) and (14), we write down Maxwell equation:
\begin{equation}
\nabla \times {\bf B} - (1/c)\left[\varepsilon(\omega) + \frac{3 \pi
q \mu^{3}(\omega)A^{2}}{mN^{2} e^{2} \omega^{2}} \right]
\frac{\partial \bf E}{\partial t}  = 0
\end{equation}
From this equation it is evident that the influence of the nonlinear
term is equivalent to the change of the dielectric permittivity or
the refraction index of the medium. When an electromagnetic wave is
propagating in this medium the refraction index is $H = c/v_{ph}$
and it depends on the wave frequency. Hence, the dispersion of the
electromagnetic radiation depends on the refraction index. From
$H^{2} = \varepsilon$ it turns out that the refraction index is
equal to $H = H_{L} + H_{NL}$, where $H_{L}^{2} =
\varepsilon(\omega)$, while $H_{NL} \approx H_{2} A^{2}$ and for
$H_{2}$ we have:
\begin{equation}
H_2 = 6 \pi q \mu^{3}(\omega)/m N^{2} e^{2} \omega^{2}
\end{equation}
Therefore, if $H_{2}>0$ the refraction index in the cavern $H =
H_{L} + H_{NL}$ turns out to be larger than the refraction index of
the ambient beyond the Debye sphere \citep{mac15}, which remains
equal to $H=H_{L}$.

Finally in the whole Debye volume let us separate  rays directed to
the observer. Due to the linear diffraction these rays has to
diverge, feature angular diffusion across the the line of
observation and before leaving the Debye volume they have to be
confined within the cone with the opening angle $2 \theta_{D}$ where
\begin{equation}
\theta_{D} \approx \frac{\lambda}{r_{D} H_{L}}
\end{equation}
where $\lambda$ is the wavelength of the electromagnetic wave.
However, when the rays leave the nonlinear medium and enter the
ambient with the refraction index $H_{L}$, the rays experience
nonlinear refraction. If the ray falls on the boundary between
nonlinear, optically more dense medium and linear, optically less
dense one and if the angle of incidence $\theta_{0} > \theta_{D}$
then all diffracted rays will undergo a total internal reflection.
We are interested in the regime when $\theta_{0} \approx \theta_{D}$
when rays assemble in a parallel beam and the observer sees enhanced
intensity of radiation \citep{mac15}. The limiting critical
incidence angle for the total internal reflection is determined by
the following equation:
\begin{equation}
cos \theta_{0} = H_{L}/(H_{L} + H_{2} A^{2})
\end{equation}
For the small value $\theta_{0}$ we find:
\begin{equation}
\theta_{0}^{2} \approx 2 (H_{2}/H_{L}) A^{2}
\end{equation}

Substituting (16) in (18) we find out that
\begin{equation}
H_{2} \simeq 1/\omega^{8}
\end{equation}
and if the condition is satisfied for certain frequencies in a given
region and at a given moment of time for other values of frequencies
it would not hold. The frequency dependence is very strongly
nonlinear, which implies that self-trapping will work only for a
very narrow frequency range.

\section{Discussion and conclusions}

In this paper we argue that the actual reason of FRB could be the
self-trapping phenomenon. This nonlinear mechanism implies that to
the radiation beam directed towards the observer additional rays are
added which, in the absence of the self-trapping, would pass beyond
the actual radiation pattern. As a result, the observer, while the
beam is being self-trapped sees an enhanced intensity of radiation
in a very narrow frequency range! This scenario is self-sustaining
and fully autonomous because unlike many other mechanisms it does
not require additional, external sources of energy. Besides,
self-trapping depends on a quite large number of parameters, in
particular, on the proper value of ratio of the wave amplitude to
the amplitude of an incident electrostatic wave, on the temperature
of the medium, direction of these waves relative to the line of
sight. Any kind of, even a slight, deviation of any of these
parameters from the ``favorable'' values may lead to the violation
of the nonlinearity condition. That is why this is a very finely
tuned, random and very rare phenomena. Its occurrence and the
arrival of the self-trapped, self-focused enhanced beam to the
observer has to be a totally random and extremely rare phenomenon.

Initially we select a volume, which, at the moment when they pass
through a randomly appearing nonlinear medium. contains waves
directed towards the observer. It can be said that rays in this
volume constitute a cylindrical beam with maximum energy
concentrated at its center. The area of maximum intensity at the
same time is optically thicker one \citep{akh68}. However the given
volume, apart from the rays directed to the observer, contains also
rays which propagate with some nonzero angle to the line of sight.
Most of these rays, providing they pass only through a linear medium
would not reach the observer. However, if these rays move from
optically thicker area to optically less thick area they would be
refracted towards the maximum energy area. Nonlinear area, selected
by us, is significantly smaller than the plasma cloud in which the
waves of the given frequency are generated. Therefore, it is
reasonable to suppose that a significant number of the given
frequency waves pass through the nonlinear region with propagation
directions constituting small angles to the line of sight, which
satisfy the condition of the total internal reflection at the
boundary between the linear and nonlinear media. Evidently the
coincidence of the channel axis with the line of sight has to be
totally random. That is why FRB are happening rarely and on a
totally random basis.

Even a slight alteration of these parameters leads to the violation
of the self-trapping condition and, therefore, disappearance of the
wave intensity enhancement - disappearance of the burst. If our
model is correct and relevant to actual FRB, the self-trapping
condition may hold only for a few milliseconds. Hence, it is
reasonable to expect that the probability of the coincidence between
the line of sight and the direction of self-trapping has to be quite
small. Therefore, what could be a serious drawback for a commonly
occurring phenomena in this case 'works' just the opposite way - it
strengthens our confidence in believing that self-trapping could be
the very reason of the appearance of this extremely rare and
energetic phenomenon - fast radio burst or FRB. Additionally,
self-trapping mechanism does not exclude other, physically
plausible, repetitive or non-repetitive, catastrophic or
non-catastrophic mechanism proposed for the FRB. Moreover, we
believe that sel-trapping may be the very 'beaming' mechanism which
might be needed for interpreting FRB as narrowly beamed radio bursts
\citep{kat17}. Any of those mechanisms giving credible explanation
of rarity of FRB, coupled with self-trapping mechanism would imply
the simultaneous occurrence of two, quite rare processes. Probably
this is the very reason why FRB are not just rare, or very rare, but
ultra-rare phenomenon, until now observed only from very distant,
extragalactic sources.

\section*{Acknowledgments}

This research was supported by Shota Rustaveli National Science
Foundation of Georgia (SRNSFG) [grant number FR/516/6-300/14].
Andria Rogava wishes to thank for hospitality Centre for
mathematical Plasma Astrophysics, KULeuven (Leuven, Belgium), where
a part of this study was finalized.

\bsp    
\label{lastpage}
\end{document}